\documentclass[peerreview, draftcls, 12pt]{IEEEtran}
\usepackage[dvips]{graphics}
\usepackage[dvips]{changebar}
\usepackage{amsmath,bm}
\usepackage{amsfonts}
\usepackage{graphicx}
\usepackage{epsf, epsfig}

\ifCLASSINFOpdf
\else
\fi

\begin{document}
%
\title{\huge Multi-Gigabits Millimetre Wave Wireless Communications for 5G: From Fixed Access to Cellular Networks}
%
%
%
\author{Peng~Wang,~\IEEEmembership{Member,~IEEE},
        Yonghui~Li,~\IEEEmembership{Senior Member,~IEEE},
        Lingyang~Song,~\IEEEmembership{Member,~IEEE},
        and~Branka~Vucetic,~\IEEEmembership{Fellow,~IEEE}
\thanks{Peng Wang, Yonghui Li and Branka Vucetic are with the School of Electrical and Information Engineering, the University of Sydney, Sydeny, Australia. E-mails: \{peng.wang, yonghui.li, branka.vucetic\}@sydney.edu.au.}
\thanks{Lingyang Song is with the School of Electronics Engineering and Computer Science, Peking University, Beijing, China. E-mail: lingyang.song@pku.edu.cn.}
}

\maketitle

\begin{abstract}
    With the formidable growth of various booming wireless communication services that require ever-increasing data throughputs, the conventional microwave band below 10 GHz, which is currently used by almost all mobile communication systems, is going to reach its saturation point within just a few years. Therefore, the attention of radio system designers has been pushed towards ever-higher segments of the frequency spectrum in a quest for capacity increase. In this article, we investigate the feasibility, advantages and challenges of future wireless communications over the E-band frequencies. We start from a brief review of the history of E-band spectrum and its light licensing policy as well as benefits/challenges. Then we introduce the propagation characteristics of E-band signals, based on which some potential fixed and mobile applications at the E-band are investigated. In particular, we analyze the achievability of non-trivial multiplexing gain in fixed point-to-point E-band links and propose an E-band mobile broadband (EMB) system as a candidate for the next generation mobile communication networks. The channelization and frame structure of the EMB system are discussed in details.
\end{abstract}


%

\section{Introduction}
%
%
%
%
    \IEEEPARstart{I}{n} recent years, video on demand, video conferencing, online gaming, e-education, and e-health are being introduced to a rapidly growing population of global subscribers such as laptops, tablets and smart phones. The formidable growth of the demand for these communication services requires ever-increasing data throughputs. To cater for this growing demand, many advanced technologies have been adopted in the current fourth-generation (4G) systems such as LTE and Mobile WiMAX to substantially increase the transmission rate. These technologies, including orthogonal frequency division multiplexing (OFDM), multiple-input multiple-output (MIMO), multi-user detection, advanced channel coding (e.g., turbo and low-density parity-check (LDPC) coding), adaptive coding and modulation, hybrid automatic repeat request (HARQ), cell splitting and heterogeneous networking, have enabled the achievable spectrum efficiency very close to the theoretical limits. Existing cellular systems are all operated below 10 GHz frequency bands that have been already heavily utilized. Therefore, there is little space to further increase the transmission rate in these frequency bands. The attention of radio systems designers has been pushed towards ever-higher segments of the frequency spectrum in a quest for capacity increase.

    The millimetre wave (MMW) band from 30 to 300 GHz offers large swathes of spectrum \cite{Lockie_MMag09}, potentially forming the basis for the next revolution in wireless communications. As predicted in \cite{Pi_CMag11}, after excluding some sub-bands with severe atmospheric absorption and assuming 40 percent of the remaining spectrum potentially being made available over time, a possible 100 GHz new spectrum among the MMW band could be opened up for future mobile communication use. This is, however, an optimistic forecast as this possible 100 GHz spectrum is discretely distributed in the overall MMW band with distinct channel characteristics and various service restrictions imposed by regulators in different countries. Uniting these discrete segments of bandwidths collectively for mobile broadband communication use will remain a great challenge. Comparatively, the frequency bands 71-76 GHz and 81-86 GHz, collectively called the E-band, have been released by the International Telecommunication Union (ITU) to provide broadband wireless services \cite{FCC03}. Different from the severe oxygen absorption in the 60 GHz band, that contributes about 15 dB/km of attenuation in addition to free-space losses, atmospheric absorption above 70 GHz drops significantly to less than 1 dB/km and rises again after 100 GHz due to molecular effects \cite{FCC97}. Therefore, the E-band opens a large frequency window with low atmospheric attenuation, making it very suitable for long-distance wireless transmissions. This 10 GHz spectrum in the E-band, which is about 50 times the bandwidth of the entire current cellular spectrum, is by far the wildest ever allocated by the Federal Communications Commission (FCC) at any one time, and can provide 5 GHz bandwidth per channel for accommodating multi-gigabits per second (Gbps) and even higher data rates with greatly reduced latency over large distances.

    There have already been some commercial E-band wireless systems for fixed point-to-point communications. For example, by utilizing the leading-edge radio frequency (RF) Monolithic Microwave Integrated Circuit (MMIC) technology \cite{Pucel_MMag12}, the E-link 1000 G1 radio from E-band Communications Corporation can provide best-in-class E-band link performance for Gbps data rates over a distance of up to a few kilometres. It has been forecast that, in the near future, the next fifth generation (5G) of cellular communication systems will be developed over untapped MMW bands. The superior propagation characteristics of E-band frequencies enable this band a preference over the other segments of MMW bands. Although E-band transceivers are presented with new design challenges such as increased phase noise, limited amplifier gain and the need for transmission line modelling of circuit components, the electronics industry develops rapidly that produces component electronics with ever reducing physical sizes and power consumption. This makes the hardware preparation for a mobile communication system over E-band going to be ready. The combination of cost-effective CMOS technology and high-gain, steerable antennas at the mobiles and base stations (BSs) will strengthen the viability of E-band communications.

    In this article, we discuss the potential of exploring the E-band spectrum for future mobile communications. We first present a brief review of the history of E-band spectrum and its light licensing policy as well as benefits/challenges. Then we introduce the propagation characteristics of E-band signals, based on which some potential fixed and mobile applications at the E-band are investigated. In particular, we analyse the achievability of non-trivial multiplexing gain in fixed point-to-point E-band links and propose an E-band mobile broadband (EMB) system as a candidate for the next generation mobile communications. The channelization and frame architecture of the EMB system are discussed in details. Finally, we conclude the article with a brief summary.



\section{E-band Spectrum}

\subsection{A Brief History of E-band}
    The E-band allocations for fixed services were first established by the ITU at the 1979 WARC-79 World Radio Communication Conference. However, not much commercial interest was shown in this band until the late 90's, when the FCC published a study on the use of the millimetre-wave bands \cite{FCC97}. Afterwards, the FCC made a historic ruling in 2002 to open up the E-band for exclusive Federal Governmental use in USA. A novel "light licensing" scheme was introduced in 2005 \cite{FCC05} and the first commercial E-band radios were installed soon after. Canada adopted the same bands with the same technical specifications and licensing regimens as USA in 2005. Also in 2005, the European Conference for Postal and Telecommunications Administrations (CEPT) released a European-wide band plan for fixed services in the E-band, which was modified later in 2009. In 2006, the European Telecommunications Standards Institute (ETSI) released technical rules for equipment operating in the E-band. Similar specifications are also effective or proposed for the United Kingdom and Australia. Nowadays, many parts of the world have followed USA and European lead, and opened up the E-band frequencies for enabling Gbps-speed point-to-point wireless transmissions.

\subsection{E-band Frequency Allocation}
    \begin{figure}[t]
        \centering
        \scalebox{1.0}{\includegraphics{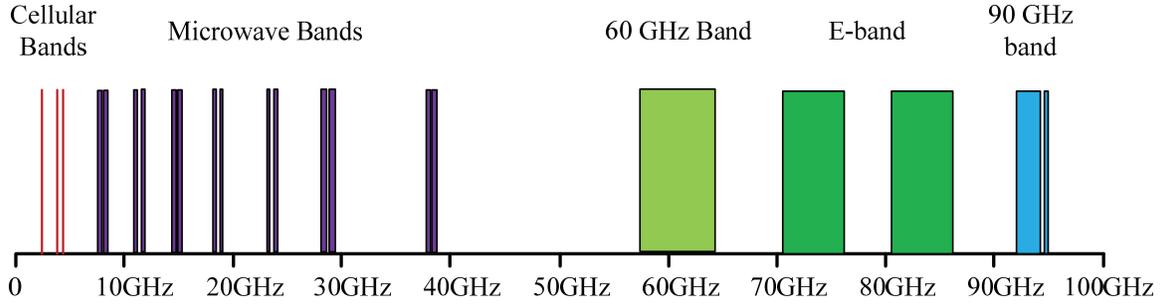}} \\
        \caption{E-band frequency allocation.}
        \label{EbandFreqAllocation}
    \end{figure}
    The E-band frequency allocation consists of the two un-channelized bands of 71-76 GHz and 81-86 GHz\footnote{In USA and Canada, the E-band spectrum also includes 92-95 GHz except 94-94.1 GHz.}, as shown in Figure \ref{EbandFreqAllocation}. Such a combined 10 GHz of spectrum is significantly larger than any other frequency allocation, enabling a whole new generation of wireless transmission to be realized. In addition, different from the lower microwave frequency bands that are sliced into sub-channels of no more than 50MHz, which in turn limits the data rate transmitted over them, the E-band spectrum is only divided into a pair of 5 GHz channels and not further partitioned. These two 5 GHz channels at E-band are 100-times the size of even the largest microwave channel. Such an un-partitioned spectrum allocation allows us to support Gbps data rates for each signal using relatively simple system architectures and modulation schemes. Radio equipment can take advantage of low order modulation modems, non-linear power amplifiers, low cost diplexers, direct conversion receivers, and many more relatively non-complex wireless building blocks, leading to reduced system cost and complexity whilst increased reliability and overall radio performance.

\subsection{Light Licensing for E-band}
    In many countries, the conventional license application of microwave bands from regulators requires a long period measured in months or even years. The corresponding licensing fee is determined by formulas depending on either data rate transmission, bandwidth use, or both. Such formulas can result in prohibitively high license fees for high capacity systems. To promote E-band commercialization, the national wireless link regulators and administrators in many countries have introduced innovative and streamlined "light licensing" \cite{E_Licensing} schemes for managing this band, enabling the E-band an attractive alternative to existing licensed frequency bands. The "light licensing" policy allows the E-band licenses to be applied for in minutes and at costs of a few tens of dollars per year, significantly faster and cheaper than traditional licensing. Thousands of fixed E-band radios have been registered and installed in these countries. The typical E-band license fees in several countries are listed in Table \ref{LightLicensing}.

    \begin{table}
        \centering
        \caption{Typical E-band license structures and license fees in some countries}
    \begin{tabular}{|c|c|c|}
        \hline
        Country     & License Structure               & License Fee\\
        \hline
        \hline
        USA         & On-line light license          & \$75 for 10-year license\\
        \hline
        UK          & Light license             & \pounds50 per year \\
        \hline
        Russia      & Light license             & Minimal registration fee\\
        \hline
        Australia   & Light license             & AUD187 per year\\
        \hline
    \end{tabular}
        \label{LightLicensing}
    \end{table}

    The "light licensing" policy comes from three unique characteristics of the E-band. First, since there are very few E-band services currently, it is argued that the spectrum at E-band frequencies is no longer scare. Second, the high frequencies at E-band allow the systems to adopt highly directional antennas and communicate via highly focused "pencil beam" transmissions, leading to dense configuration of communication links without interference concern and thus a high degree of frequency reuse. Third, the E-band frequencies are configured as a single pair of 5GHz channels, which makes the traditional frequency planning/coordination unnecessary and the related interference analysis significantly simplified. Thus the E-band administration and cost of license are dramatically reduced. The "light licensing" at the E-band reflects the ease of coordinating, registering and licensing, and sets license fees that cover administrative costs, but does not penalize the high data rates and bandwidths that are required for ultra-broadband services. It is worth noting that despite the name "light licensing", the possession of such a license still gives the link operator the same full benefits of a traditional link license, including link registration, "first come first served" rights and full interference protection.

\subsection{Benefit of E-band over Other Wireless Technologies}
    There are many technologies available to provide wireless broadband connectivity and fiber-like services. These technologies include WiFi, 60 GHz wireless, free space optics (FSO) and so on. E-band wireless systems offer significant benefits over them with the following advantages \cite{E_Benefit}:
    \begin{itemize}
        \item[$\bullet$] High antenna gains and allowable output power: Thanks to the small wavelength of E-band signals, it is possible to realize large gains from relatively small antennas at E-band frequencies. In addition, the FCC permits E-band radios to operate with up to 3W of output power, significantly higher than available at other millimetre-wave bands (for example, 25 dB higher than the 10 mW limit at 60 GHz). The high antenna gain and high output power allow E-band radios to overcome the higher rain fading and foliage losses experienced at E-band frequencies.
        \item[$\bullet$] Guaranteed high data rates: E-band offers much higher data rates, e.g., Gbps and above, than any other wireless technology. Such high rates are guaranteed even under deteriorated transmission conditions such as rain, which beats WiFi, WiMAX and other broad-coverage technologies whose system performance depends heavily on the radio and user environments;
        \item[$\bullet$] Long distance transmissions with robust weather resilience: E-band wireless allows Gbps-level transmission over a very long distance up to 12 miles, much longer than those supporting similar data rate such as 60 GHz and FSO systems. This long-distance transmission is robust to almost all environmental conditions such as fog, dust, air turbulence and other atmospheric impairment that can disable optical links for hours;
        \item[$\bullet$] Low-cost and rapid licensing policy providing guaranteed interference protection: Under the "light licensing" policy, licenses for E-band links can be obtained much faster and cheaper than those for traditional microwave bands, and in the meanwhile provide the full benefits of traditional link licenses that grant full interference protection from other nearby wireless sources. Even in the unlikely event of interference, the full weight of the wireless regulator is available to identify and remove the interference source;
        \item[$\bullet$] Cost effective, fiber-like wireless solution: The cost of high capacity E-band wireless systems is only a fraction of that of the buried fiber alternatives. Installed wireless systems have payback periods of months when compared to the costs of trenching new fiber. Installing dedicated wireless technology can often be more economic than leasing fiber-provided high capacity services.
    \end{itemize}

    A summary of the most important system parameters and network characteristics of various broadband techniques are detailed in Table \ref{Eband_Benefit}.

    \begin{table}
        \centering
        \caption{Comparison of different broadband techniques}
    \begin{tabular}{|c|c|c|c|c|c|c|}
        \hline
                            & WiFi              & 3/4G                  & 60GHz             & FSO           & Fiber         & E-band \\
        \hline
        \hline
        Data rate           & about 1 Mbps,     & about 10 Mbps,        & 100$\sim$1000     & 100$\sim$1000 & up to 100s    & mutiple \\
                            & unstable          & unstable              &  Mbps             & Mbps          & of Gbps       & Gbps \\
        \hline
        Transmission        & 20 yards          & 2 miles               & 500 yards         & 200 yards     & up to         & up to \\
        distance            &                   &                       &                   &               & 60 miles      & 12 miles \\
        \hline
        Licensing           & free for          & licensed, spectrum    & free for          & not           & N/A           & light \\
                            & unlicensed use    & very scare            & unlicensed use    & regulated     &               & licensing \\
        \hline
        Licensing cost      & N/A               & high                  & N/A               & N/A           & N/A           & low \\
        \hline
        Licensing           & N/A               & months/years          & N/A               & N/A           & N/A           & minutes/hours \\
        application period  &                   &                       &                   &               &               &    \\
        \hline
        Guaranteed inter-   & No                & Yes                   & No                & No            & Yes           & Yes \\
        ference protection  &                   &                       &                   &               &               &    \\
        \hline
        Installation time   & hours             & months/years          & hours/days        & hours/days    & months/years  & hours/days \\
        \hline
        Installation cost   & low               & high                  & medium            & medium        & high          & medium \\
        \hline
    \end{tabular}
        \label{Eband_Benefit}
    \end{table}

\subsection{Technical Research Challenges on the E-band Communications}
    Though numerous benefits are presented above, there are still some challenging technical issues that must be addressed before the commercialization of the E-band frequencies. They include 1) severe E-band propagation loss; 2) unclear channel modelling at such high frequencies; 3) high transceiver complexity in such a large MIMO systems with a massive number of antennas and 4) new transceiver design due to the hardware constraint at E-band transmitters/receivers - a large number of antennas are driven by a limit number of radio-frequency (RF) chain due to the high cost and power consumption of the latter. However, as to be elaborated in the remaining part of this article, all these issues can potentially be, and are already being effectively addressed. The severe propagation loss can be readily compensated through deploying a large number of transmit/receive antennas that provide significant beamforming gains. Several research groups have already conducted E-band propagation measurements in real urban environments \cite{Rappaport_ICC14}, providing some fundamental hints for the proper modelling of E-band channels. Some initial and efficient channel estimation algorithms \cite{Heath_JSTSP14} that utilize the channel sparsity and hybrid precoder design \cite{Heath_TWC13} that relieves the high hardware costs have also been proposed. All these delightful progresses have enabled E-band a very promising candidate frequency segment for future 5G wireless broadband mobile communications.

\section{E-band Propagation}
    While signals at lower frequency bands propagate for several tens of miles and penetrate easily through buildings, E-band signals can travel only a few miles or less and do not penetrate solid materials very well. However, these characteristics are not necessarily disadvantageous. In fact, the propagation loss can be exploited to reduce interference, increase frequency reuse and prevent eavesdropping, thus providing very efficient spectrum utilization and increasing security of communication transmissions.

\subsection{Free Space Propagation}
    Due to the small wavelength of E-band signals, transmissions over the E-band are principally contributed by line-of-sight (LoS) components. According to the free space transmission model, the path gain of the LoS link between two omnidirectional antennas with distance $D$ is mathematically expressed as
    \begin{equation}
        G = G_T G_R\frac{\lambda^2}{(4 \pi D)^2}
        \label{FreeSpacePathLoss}
    \end{equation}
    where $G_T$ and $G_R$ are, respectively, the gains of transmit and receive antennas and $\lambda$ is the signal wavelength. It is seen from (\ref{FreeSpacePathLoss}) that, given $G_T$, $G_R$ and $D$, the path gain is proportional to $\lambda^2$, indicating that the E-band transmissions suffer much more power loss than those over conventional microwave bands. For example, the propagation at 75GHz is 30dB worse than that at 2.4GHz (the operating frequency for WiFi networks). Thus to guarantee the same signal power (and in turn the same quality of service) at the receiver, the transmitted power at 75GHz must be 30dB higher than that at 2.4GHz. This makes the signal transmission/reception through a single omnidirectional antenna practically infeasible in E-band systems.

    One approach to compensate the severe E-band power loss is to equip a massive number of antennas at both link ends to provide a large beamforming gain. Different from conventional microwave systems where the large-size antennas must be sufficiently spaced and may lead to an extraordinarily large transmitter/receiver aperture sizes, this approach can be easily implemented in E-band systems as the antenna size and spacing scale down with the wavelength. The synthesized low-cost antenna arrays can be electronically steered to provide adaptive yet highly directional links permitting a flexible deployment. In principle, the number of antenna elements that can be packed into a given aperture size is increased by four times for every doubling the operating frequency, providing about 6dB beamforming gain at each link end if these antennas are compactly located to form an equivalent directional antenna for steering a "pencil beam". When the beamforming gains at both link ends are taken into consideration, the overall power gain of the link then scales as $1/\lambda^2$. Therefore, the propagation at 75GHz becomes 30dB better than that at 2.4GHz, which implies a significant redeeming feature of multiple antenna transmissions in the E-band.

\subsection{Blockage, Multi-Path and Scattering}
    Pure free space propagation between the transmitter and receiver happens only when the LoS component is present and no building/obstacle is available around. In practice, an E-band communication link is always located within a building group area, in which the buildings, cars and even human beings may either block the LoS transmission or "bend" the signal impinged on their surfaces. The corresponding propagation characterizations of E-band signals are very different from that of traditional microwave ones. Due to the small wavelength at the order of several millimetres, transmissions over E-band are effectively blocked by obstacles such as wooden boards and brick walls. In addition, E-band signals are not prone to diffraction either when encountering an obstacle, which is similar to light waves. Reflection constitutes the most received signal power among all non-LoS (NLoS) links. Principally, the signal power received from each reflected link may be quite lower than that from a LoS link. This is because, besides partial absorption by reflecting materials, E-band signals encounter greater diffusion and less specular reflection than microwave signals due to the relatively "rougher" reflecting material surface compared to their signal wavelengths. Even though, it has been experimentally validated that \cite{Rappaport_ICC14} these NLoS links can still provide substantial link connection and coverage extension in MMW cellular systems, especially when LoS transmissions are unavailable. According to the E-band propagation measurements conducted by NYU WIRELESS in the dense urban environment of New York City \cite{Rappaport_ICC14}, the path loss exponent for NLoS propagation is 5.88 with a shadow factor of 14.19 dB, which is a result of several different paths of great dynamic range supported over a wide range of angles. Therefore, though the large buildings on every city block and crowed streets cause numerous blockages, they also create reflections and scatters between the transmitter and receiver with slight more path loss and fewer multi-path components than those measured at a lower frequency of 28 GHz. This indicates that E-band transmissions will be able to rely on multi-path environments and directional antennas to overcome additional propagation loss at E-band. In addition, the less number of multi-path components relative to those of the transmit/receive antennas endue a sparse nature of the E-band propagation channel, which may significantly reduce the operational complexity involved in the channel estimation and transceiver design. Provided that the angle of departure (AoD) and arrival (AoA) information of each path is available, we can combine multi-path components with different AoDs/AoAs to significantly improve the path loss exponents and link margins through beamforming and beam combining. This will make it feasible to deploy a mobile communication network over E-band with reasonable BS coverage and acceptable outage performance.

\subsection{Other Attenuation Factors at E-band}
    In addition to the power loss during the free space and reflected/scattered propagations, the transmission over E-band also suffers from some other attenuation factors, as detailed below.

    \emph{Atmospheric attenuation}: When traveling through the atmosphere, the E-band signals may be absorbed by molecules of oxygen, water vapour and other gaseous atmospheric constituents. Fortunately, these losses are merely about 0.5dB/km in total, much less than those at 60GHzs and 100GHz above and close to that of the popular microwave frequencies. This makes the E-band frequencies very favourable for ratio transmissions over many miles under clear conditions.

    \emph{Fog and clouds}: Since the fog and cloud particles are much smaller than the E-band wavelengths, the attenuation caused by them is almost negligible, e.g., only an attenuation of 0.4dB/km is led by thick fog at density of 0.1g/m3 (about a visibility of 50 meters). Comparatively, the attenuation for an FSO optical signal caused by heavy fog could be about 200dB/km, due to the similar magnitudes of the signal wavelength and fog/cloud particles.

    \emph{Dust and other small particles}: Similar to fog and cloud particles, the magnitudes of these particles are much smaller than the E-band wavelengths, making them essentially invisible to E-band transmissions.

    \emph{Rain}: Transmissions at E-band experience significant attenuation in the presence of rain \cite{ITU_R05}, which places practical limits on the link distances. For example, "heavy" rainfall at the rate of 25mm/hour can lead to over 10dB/km attenuation at E-band frequencies. The corresponding attenuation even reaches up to 30 dB/km in the case of tropical rainfall with rate 100mm/hour. Fortunately, most intensive rain tends to fall in limited parts of the world, mainly the equatorial countries. In other countries such as United States, Canada and Australia, such severe weather generally occurs only in very short bursts. It tends to fall in small and dense clusters within a larger and lower intensity rain cloud, and is usually associated with a severe weather event that moves quickly across the link path. Therefore, rain outage tends to be short and is only problematic on longer distance transmissions. With well-understood information of rainfall characteristics in particular regions, it is easy to design E-band ratio links capable of overcoming the worst weather conditions via adaptive transmit power control, or predict the levels of weather outage of longer links.

    \emph{Ice crystals and snow}: Ice crystals and snow do not cause appreciable attenuation, even if the rate of fall exceeds 125mm/h. This is due to the much-reduced loss of ice compared to water.

    \emph{Foliage}: Foliage losses are significant at E-band frequencies and may be a limiting propagation impairment for E-band transmissions. For example, the foliage loss at 75GHz for a penetration of 8 meters (roughly equal to the diameter of a large tree) is about 20dB.

    In summary, E-band propagations exhibit comparable characteristics to those at the widely used microwave bands, and with well characterized weather characteristics allowing rain fade to be understood, link distances of several miles can confidently be realized.

\section{Fixed E-band Applications}
    A wide range of fixed services are realizable over E-band frequencies. The following are some examples.

    \emph{Last-mile access}: In many communities, the last-mile access technique represents a major remaining challenge because the cost of providing high-speed, high-bandwidth services to individual subscribers in remote areas can be higher than the service provider would like. Laying wire and fiber optic cables is an expensive undertaking that can be environmentally demanding and require high maintenance. Many experts believe that broadband wireless networks will eventually solve this difficulty and meet everyone's needs. E-band frequencies provide a promising solution in terms of flexibility, speed and cost of construction.

    \emph{Wireless backhaul}: With the rapid growth of mobile data traffic, traditional backhaul that utilizes narrow bandwidth is consequently regarded as a potential bottleneck for the overall cellular system. E-band offers a cost-effective and flexible alternative to fiber for future backhaul. Access points and BSs can be easily connected via E-band links, providing Gbps backhaul capacity to solve this bottleneck problem.

    \emph{Network recovery}: In the case of fiber breakage, a fixed point-to-point E-band link can be used to provide the temporary service restoration, owing to its much shorter set-up time and lower cost in comparison with those required to restore the original fiber link.

    \emph{Campus LAN}: Fixed E-band links can also be installed to directly build up a gigabit wireless LANs within a building group (e.g., campus) as the extension of fiber-optic communication networks. High speed gigabit access will be maintained within both the wireless and wired parts of the overall communication networks, but without problems and expenses related to fiber installation.

    \emph{Storage access}: Machine to machine connectivity for the storage area networks could be easily established via fixed point-to-point E-band links with excellent data security and high availability.

    In all the above applications, both terminals of a communication link are usually fixedly located, e.g., on the tops or side walls of buildings such that LoS transmissions are guaranteed. Therefore, the corresponding channel is mainly contributed by LoS transmissions, and other effects such as multi-path, foliage loss and atmospheric stratification are not significant due to the extremely narrow beams in which the radiation propagates. The primary sources of link impairments come from adjacent link interference, rain attenuation and antenna perturbation. The adjacent link interference occurs when the LoS power of one link is directed into the main lobe or a side-lobe of the receive antennas in adjacent links. This impairment can be avoided in advance via proper organization of link deployments. The rain attenuation has been well understood and the resultant impairment can be compensated via adaptive transmit power control. The antenna perturbation, potentially incurred by wind-induced pole sway or other environmental concerns, may lead to severe mismatch between the directivities of transmit/receive antennas. Some robust and computationally efficient beam alignment technique may be required to combat this problem.

    A main concern for fixed E-band systems is whether a high-capacity link can be guaranteed between the transmitter and receiver to support gigabits or even tens-of-gigabits throughputs over a given distance. As aforementioned, multiple antenna techniques are essential in the E-band to provide beamforming gain for compensating the severe propagation loss. To further enhance the link capacity, we need rely on multiple antenna techniques to achieve a multiplexing gain such that transmissions of multiple spatially independent signal streams can be supported simultaneously without interfering with each other. This is, however, not a trivial problem in the E-band systems. Unlike rich-scattering microwave channels where independent Rayleigh faded coefficients between different antenna links enable the achievable multiplexing gain to increase linearly with the minimum number of transmit/receive antennas, the fixed E-band channels are dominated by nearly deterministic LoS components and the achievable multiplexing gain heavily depends on the antenna deployments at both link ends.

    In spite of this, it has been shown \cite{Gesbert_TC02}\cite{Oien_TWC07} that the maximum multiplexing gain is still principally achievable in fixed E-band channels, provided that the geometrical distributions of the antennas at both link ends are carefully designed. For example, in an E-band LoS MIMO channel with aligned uniform linear antenna arrays (ULAs) at both the transmitter and receiver, the maximum multiplexing gain is achieved when the following Rayleigh distance criterion is fulfilled \cite{Gesbert_TC02}.

    \begin{equation}
        D = D_{Ray} = \frac{\max\{N_t, N_r\}d_td_r}{\lambda}
        \label{DRayleigh}
    \end{equation}
    where $N_t (N_r)$ and $d_t (d_r)$ are, respectively, the number and spacing of the transmit (receive) antennas. In this case, the channel contains $\min\{N_t, N_r\}$ eigenmodes with equal channel gains, indicating that the maximum multiplexing gain, $\min\{N_t, N_r\}$, is indeed achievable and thus that many spatially independent signal streams can be supported simultaneously. Similar observations have also been made in the general situation when the ULAs at both ends have arbitrary orientations \cite{Oien_TWC07}. However, the antenna spacings, $d_t$ and $d_r$, in a practical E-band LoS MIMO system may be limited by the physical sizes of the transmitter/receiver and cannot be arbitrarily large. Consequently, the communication distance that satisfies the Rayleigh distance criterion is also limited, indicating that the maximum multiplexing gain is not always achievable in practice.

    \begin{figure}[t]
        \centering
        \scalebox{1.0}{\includegraphics{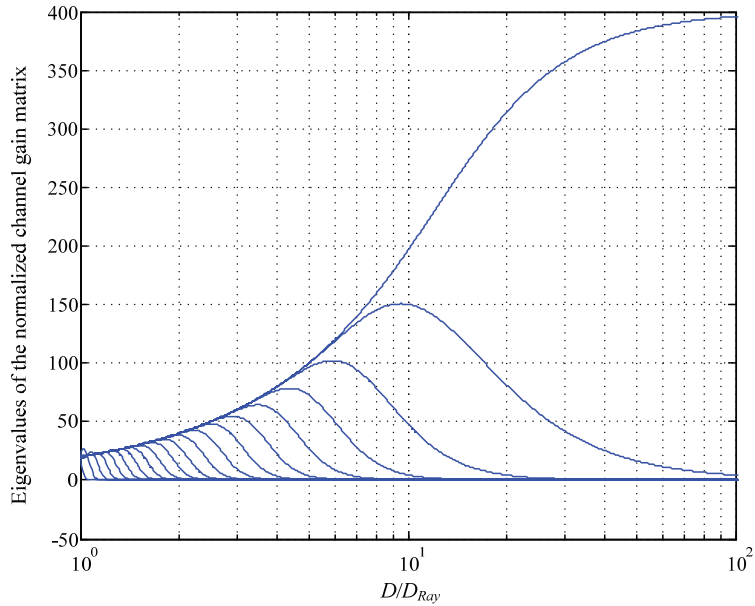}} \\
        \caption{The eigenvalue curves for the normalized channel gain matrix of a ULA-based E-band LoS MIMO channel with 20 antennas at both the transmitter and receiver, i.e., $N_t = N_r = 20$.}
        \label{EigenvalueCurves}
    \end{figure}

    Fig. \ref{EigenvalueCurves} shows all the eigenvalue curves of an aligned ULA-based E-band LoS MIMO channel with 20 antennas at both ends, i.e., $N_t = N_r = 20$. In Fig. \ref{EigenvalueCurves}, we assume a far-field distance between the two link ends such that the channel coefficients for all antenna links have approximately the same amplitude that is normalized to one for convenience. It is seen that, for the E-band system beyond the Rayleigh distance (i.e., $D > D_{Ray}$), although the channel may still be of full rank, some of its eigenmodes are very poor and signal transmissions over them will be very inefficient in practice. For convenience, we denote by $\mu_m(D)$  the $m$-th largest eigenvalue of a ULA-based E-band LoS MIMO channel gain matrix and account the $m$-th eigenmode as an effective eigenmode if $\mu_m(D)/\mu_1(D) \geq \gamma$, where $\gamma$ is a threshold related to the system working signal-to-noise ratios (SNRs). As a consequence, the number of effective eigenmodes can be referred to as the effective degree of freedom (EDOF) of the channel. It is shown in \cite{WP_TWC14} that, when $N_t$ and $N_r$ are sufficiently large, the farthest distance that can provide an EDOF of $m$ (and in turn can support $m$ spatially independent signal streams) is mathematically given by
    \begin{equation}
        D_{\max}^{(m)} = c_m(\gamma)\frac{N_td_tN_rd_r}{\lambda} \approx c_m(\gamma)\frac{(N_t-1)d_t(N_r-1)d_r}{\lambda} = c_m(\gamma)\frac{D_tD_r}{\lambda}
        \label{Dmax}
    \end{equation}
    where $D_{\max}^{(m)}$  is referred to as the effective multiplexing distance of EDOF-$m$, $D_t = (N_t-1)d_t$ and $D_r = (N_r-1)d_r$ are the aperture sizes of the transmit/receive ULAs, respectively, and $c_m(\gamma)$ is a constant function. Equation (\ref{Dmax}) indicates that, the farthest distance that can support a given number of spatially independent signal streams at a finite SNR is mainly determined by the product of the aperture sizes of the transmit/receive ULAs, instead of the numbers of antennas at both ends. Hence to support a more number of spatially independent signal streams in a ULA-based E-band LoS MIMO channel, we must either increase the product of transmit/receive aperture sizes or reduce the communication distance.

\section{E-band Mobile Broadband (EMB) Communications}
    In this section, we discuss the feasibility and challenge of establishing an E-band mobile broadband (EMB) network. As mentioned earlier, E-band signals do not penetrate solid materials very well. This implies that the overall EMB networks can be effectively isolated into indoor and outdoor networks by brick walls of the buildings. For indoor networks, plenty of reflecting materials are present, making NLoS transmissions (also named as diffuse links) very common in such scenarios. Therefore, indoor mobile users can easily get access to the network via access points installed in each room without suffering from weather impairment. The Doppler effect is not a concern either as the relatively small indoor serving area restricts the user mobility. Since this scenario has been extensively investigated for communication over other segments of MMW bands such as 60 GHz, we will not discuss it here. In what follows, we assume that handoff between indoor and outdoor networks is guaranteed via the access points equipped at the entrances of the buildings and mainly focus on the outdoor networks.

    A common myth in the wireless engineering community is that rain and foliage attenuation make E-band spectrum practically useless for outdoor mobile communications. However, the outdoor EMB network can overcome these issues and provide seamless user experience after adopting the following potential techniques.

\subsection{Dense EMB BS Deployment}
    To guarantee a reasonably high successful link connection probability between the BSs and mobile users and provide sufficiently good coverage, it is preferable to equip BSs densely in a given EMB network area so as to combat both the severe path attenuation experienced by E-band signals and the possible block of LoS transmissions caused by surrounding buildings/obstacles. The BS antennas could be located adaptively according to the topography and architectural construction of the severing area, e.g., on the surface of buildings or the top of lampposts along the streets and at each street corner. The E-band propagation measurements conducted by NYU WIRELESS in the dense urban environment of New York City have revealed that \cite{Rappaport_ICC14}, for inter-site distances up to 200 meters, atmospheric attenuation is of a negligible degree and the rain attenuation is only about 2 dB for a heavy rainfall of 25 mm/hr. Therefore, a cell size on the order of about 200 meters, similar to today's microcell sizes, is sufficient to guarantee qualified LoS links in urban environments. Thanks to the distinctive narrow beam technique adopted at E-band, the interference among adjacent EMB BSs can be significantly supressed and thus their coverage areas can be largely overlapped.

\subsection{Adaptive Beamforming}
    Beamforming is another efficient technique to overcome the path attenuation. At the transmitter, the signal is emitted from different antennas with different phases and amplitudes, creating constructive or destructive patterns at intended or undesired receivers. At the receiver, signals from different receive antennas are combined together using a set of weight coefficients such that the power or signal-to-noise ratio (SNR) of the collected signal after combination is maximized. When a LoS link is available between the mobile user and the BS, proper beamforming/combining patterns that point to each other can be generated at both link ends so as to significantly enhance the link quality.  On the other hand when a LoS link is unavailable, adaptive beamforming is still capable of enhancing the NLoS link quality by exploiting multi-path in urban environments. In this case, the surrounding buildings, especially those with smooth surfaces made of glass or marble, could provide stronger reflection and less diffusion. Thus the signal transmission and reception can be directed to such strong reflected NLoS links using adaptive beamforming. Satisfactory link quality can still be achieved together with proper adaptive transmit power control.

\subsection{Sparse Channel Estimation}
    To explore the potential benefit of adaptive beamforming, accurate and timely channel state information (CSI) is crucial in an EMB network. Recall that a massive number of antennas are necessarily required at one or both link ends so as to provide sufficient power gain in compensating the severe E-band propagation loss. This indicates a significant increase of channel state information (CSI) overhead to be estimated at the receiver and fed back to the transmitter. Fortunately, recent research results \cite{Rappaport_ICC14} have revealed that, due to the much higher E-band signal frequencies, an E-band channel generally consists a much less number of paths between the transmitter and receiver compared with its antenna numbers equipped at both link ends, even in the dense urban environment. This indicates that an E-band channel can exhibit a sparse nature after being converted into the beam-space domain, and by utilizing this sparse property, the CSI overhead in an EMB network can be significantly reduced. A channel estimation algorithm that explores this sparsity in EMB networks has already been proposed \cite{Heath_JSTSP14}, which directly works on the sparse version of the channel matrix after been converted into the beam-space domain and can quickly estimate the AoDs/AoA and fading coefficient of each path in a bi-section searching manner. More efficient and advanced channel estimation approaches are also under investigation.

\subsection{Hybrid Transceiver Design}
    Due to the user mobility, the beamforming/combining vectors need be adaptively adjusted so that the beams are always pointing to each other as the mobile user moves. However, different from the adaptive beamforming technique for traditional microwave that can be implemented digitally at baseband, the adaptive beamforming design in E-band is restricted by the hardware constraint, where a large number of antennas are driven by a limit number of radio-frequency (RF) chain due to the high cost and power consumption of the latter. Hybrid digital-and-analogue precoder/combiner design is a practical solution to this difficulty \cite{Heath_TWC13}. With hybrid precoding/combining, the transmitters/receivers are able to apply high-dimensional RF precoder, implemented via analog phase shifters, followed by low-dimensional digital precoder that can be implemented at baseband. Near optimal unconstrained performance can be achieved at practically low cost.

\subsection{User Cooperation}
    When the surrounding buildings of an E-band network have relative rough surfaces, the reflected signal power may be much reduced, and the link quality cannot be guaranteed if the LoS link between the mobile user and BS is blocked. User cooperation may provide a solution to this situation. Specifically, we can build up certain rewarding mechanism to encourage the vacant users with good-quality links to BSs to serve as relays and help forwarding data for other users with bad link qualities. The overall network may work as follows. Firstly, all the EMB BSs continuously broadcast their pilot signals selected from a pilot set $\mathcal{P}_1$ through a signalling channel. The pilot signals used by different BSs are referred to as level-1 pilots and assumed to be mutually orthogonal. Each mobile user in the serving area, whenever it has data to transmit/receive or not, estimates the qualities of the links to different BSs based on the received level-1 pilot signals. These mobile users are then classified into directly-served (DS) users and indirectly-served (IS) users, according to the link qualities to the surrounding BSs. A DS user refers to the mobile user that has at least one BS to which the link quality is better than a certain threshold. Contrarily, an IS user refers to the user whose link qualities to all BSs are below the threshold. The operations for DS and IS users are different and introduced separately below.

    \emph{Operation for DS Users}: Each DS user chooses the BS with the best link quality as its severing BS and registers to its severing BS for future data transmission, reception or forwarding. If a DS user has data to transmit/receive, a traffic channel will be assigned by the severing BS for performing data transmission/reception in a similar way to the operation in traditional cellular networks. Otherwise if this DS user is vacant, it will keep listening to the channel and, in the meanwhile, broadcast the link quality information between it and the severing BS together with a pilot signal (referred to level-2 pilot) selected from another pilot set $\mathcal{P}_2$. By this means and assuming that all the pilots in $\mathcal{P}_1$ and $\mathcal{P}_2$ are mutually orthogonal, these vacant DS users will serve as potential relays to help the data forwarding of IS users.

    \emph{Operation for IS users}: After failing to connect to any BS due to lack of both LoS and strong NLoS reflected links, an IS user will measure the qualities of the links to all the surrounding vacant DS users who are broadcasting level-2 pilots, based on which the overall qualities of all the possible IS user-vacant DS user-BS links are calculated. Afterwards, this IS user sends a request to the vacant DS user through which the overall two-hop link quality is the best. Upon receiving and accepting this request, the selected vacant SD user then builds up an indirect link between its severing BS and the served IS user, helping the latter to register to the former indirectly. When data transmission/reception is required by this IS user, the selected DS user will apply for a traffic channel for him from its severing BS, enabling the data of this IS user to be forwarded to/from the core network.

\subsection{Hybrid EMB and 4G System}
    Due to the nature of E-band signal propagation, it is possible that, even with the abovementioned techniques, some dead spots still exist at which the mobile users cannot be served by the EMB network. Therefore, a mechanism that supports emergency communications when the communications over the E-band are not successful should be considered as part of the EMB system design. A hybrid EMB and the current 4G cellular network infrastructure may be adopted to provide better coverage and seamless user experience, while in the meanwhile preserve the benefit of gigabit transmissions at E-band. In such a hybrid network, the overall E-band frequencies are mostly utilized for data transmission. The system information, control channel and feedback channel are implemented over the current 4G frequency band. In addition, a part of 4G cellular frequencies should also be preserved for data transmission of the mobile user at dead spots.

\section{Channelization and Frame Structure}
    Although ITU has released the E-band frequencies for fixed and mobile services, there has been no specific recommendation regarding the use of the E-band or sharing arrangements with other services. Nevertheless, many countries have issued their E-band channelization plans to promote its commercialization. Fig. \ref{Channelization} below summarizes the E-band channelization plans in several representative areas. In the United States and Canada, both the 71-76 GHz and 81-86 GHz bands are divided into four unpaired 1.25 GHz segments (eight in total) without mandating specific channels within them and these segments may be aggregated without limit. In Europe, UK and Australia, a 125MHz guard band is set at the top and bottom of each 5GHz sub-band of the E-band spectrum to prevent potential interference to and from adjacent bands. In particular, UK and Australia have no explicit channel plan for the rest segments of the E-band, while Europe further divides each of the two 4.75 GHz bands into nineteen 250 MHz channels and allows aggregation of any number of channels from 1 to 19. Furthermore, the specified channels may be used for either time division duplex (TDD) or frequency division duplex (FDD) systems either within the single band or in combination with other bands.

    \begin{figure}[t]
        \centering
        \scalebox{1.0}{\includegraphics{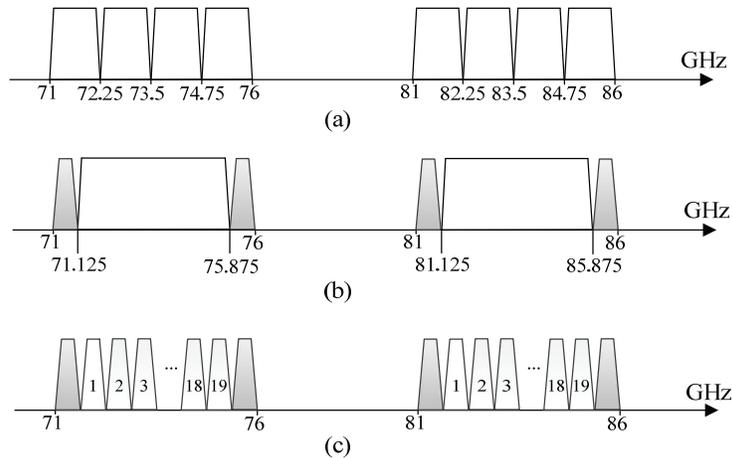}} \\
        \caption{E-band channelization in (a) the United States and Canada, (b) UK and Australia, and (c) Europe.}
        \label{Channelization}
    \end{figure}

\begin{figure}[t]
        \centering
        \scalebox{1.0}{\includegraphics{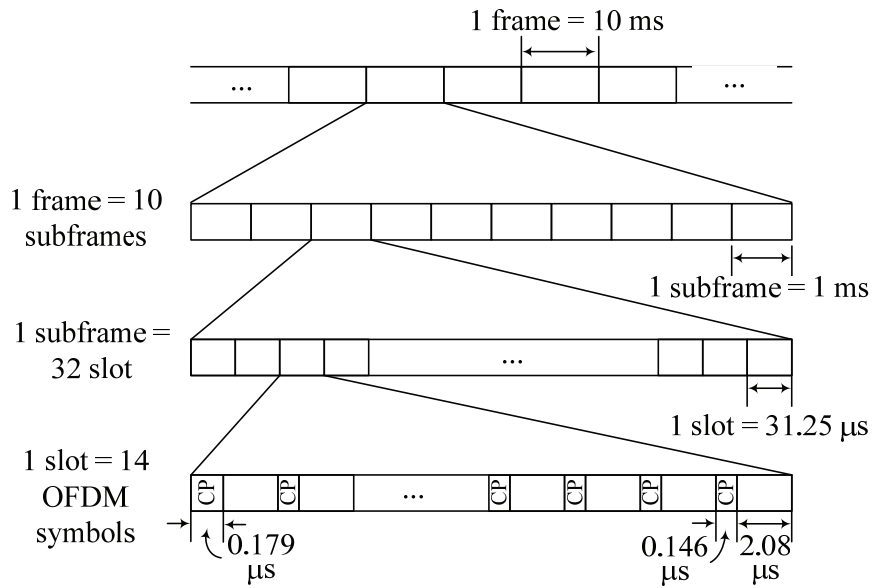}} \\
        \caption{Frame structure of the EMB system.}
        \label{FrameStructure}
    \end{figure}

    Here we propose a possible frame structure for the EMB system based on the Europe channelization plan. Note that since the Europe channelization plan is compatible with that of UK and Australia, our proposed frame structure is also applicable to the latter two countries. Following the current 4G systems, we choose OFDM as the multiplexing scheme for EMB due to its superiority in efficient multiple access and simpler equalization at the receiver. As shown in Fig. \ref{FrameStructure}, the durations of one frame and sub-frame are chosen to be 10ms and 1ms, respectively, which are the same as those of LTE systems so as to facilitate the hybrid EMB and 4G operations.

    The other parameters in the OFDM numerology are designed as follows. Since the bandwidth of each channel in the Europe channelization plan is 250MHz, we choose the sampling rate as 30.72 MHz $\times$  8 = 245.76 MHz, where 30.72 MHz is a popular frequency at which a good trade-off can be achieved between clock accuracy and cost. In addition, we select the subcarrier spacing to be 480 KHz based on the following reasons.

    \begin{itemize}
        \item[$\bullet$]	First, from the implementation viewpoint, the fast/inverse fast Fourier transform (FFT/IFFT) size, denoted by $K$, is typically a power of 2, meaning that the subcarrier spacing should have a form of $30.72 \times 2^k$ MHz for some integer $k$. The value of 480KHz satisfies this form when $k=8$;
        \item[$\bullet$]    Second, due to the high directional transmission characteristic of EMB, the corresponding maximum delay spread may be limited to a few nano-seconds, which in turn leads to a much wider coherent bandwidth than that in LTE. In this case, the subcarrier spacing of 480KHz is small enough to stay within the coherent bandwidth of most situations in EMB.
        \item[$\bullet$]    Third, by assuming that the moving speed of mobile users is no more than 120km/h, the resultant Doppler shift, $f_d$, is at most 120km/h $\times$  86 GHz/(3$\times 10^8$m/s) $\approx$ 10KHz. This value is much less than 480KHz and thus can keep inter-carrier interference (ICI) due to Doppler sufficiently low;
        \item[$\bullet$]    Forth, with a reasonable clock accuracy of 10 ppm, the corresponding clock drift at E-band is at most 10ppm $\times$ 86 GHz $=$ 860 KHz, which should be less than 2 times of the subcarrier spacing so as to enable simple system synchronization and acquisition;
        \item[$\bullet$]    Finally, the 480 KHz subcarrier bandwidth indicates an FFT/IFFT size of 512 points for the overall 250 MHz bandwidth of each channel, which is small enough from complexity concern because this size takes about $20\%$ of the RX digital baseband complexity.
    \end{itemize}

    Furthermore, since the channel coherent time is $T_c = 1/f_d \approx 0.1$ ms determined by the above calculated  Doppler shift, we divide each sub-frame into 32 slots such that each slot has a duration of 31.25 $\mu$s that is less than the channel coherent time. The number of OFDM symbols in each slot is set to be 14 with the corresponding cyclic prefix (CP) lengths being about 0.179 $\mu$s (44 samples) for the first OFDM symbol and 0.146 $\mu$s (36 samples) for the remaining 13 OFDM symbols. Such a design leads to a CP overhead of about 6.7$\%$ and provides sufficient margin to cope with the maximum delay spread and synchronization error.

\section{Conclusions}
    In this article, we have introduced the background and propagation characteristics of E-band transmissions. The potential of exploring the E-band spectrum for mobile broadband communications in the coming few decades is discussed in particular. E-band transmissions heavily rely on directional beamforming with very narrow beam widths, allowing effective suppression of interference among adjacent E-band mobile broadband BSs and significant overlap of the coverage areas of them. Also because of directional beamforming, a key challenge in the E-band mobile broadband network is to guarantee a good coverage of the overall network, especially when some mobile users do not have LoS links to the surrounding BSs. Several techniques have been discussed that can potentially solve the coverage problem and provide good link qualities regardless of the locations of the mobile users in the network area. A hybrid EMB and 4G system may provide a good tradeoff between the coverage and data rate.


%




\section*{Acknowledgment}
    The authors would like to thank Mr. Yongping Zhang, Ms. Jiahui Qiu and Ms. Miao Wang for enlightening discussions on the design of EMB frame structure.

\ifCLASSOPTIONcaptionsoff
  \newpage
\fi



%

%


\begin{IEEEbiography}[{\includegraphics[width=1in,height=1.25in,clip,keepaspectratio]{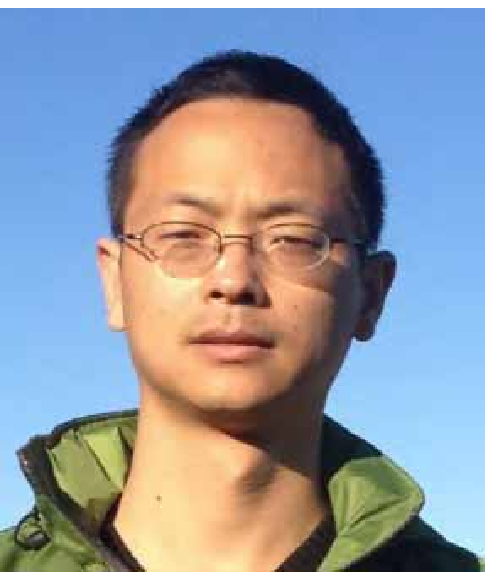}}]{Peng Wang} (S'05-M'10)
received the B. Eng. degree in telecommunication engineering and M. Eng. degree in information engineering, from Xidian University, Xi'an, China, in 2001 and 2004, respectively, and the Ph.D. degree in electronic engineering from the City University of Hong Kong, Hong Kong SAR, in 2010. He was a Research Fellow with the City University of Hong Kong and a visiting Post-Doctor Research Fellow with the Chinese University of Hong Kong, Hong Kong SAR, both from 2010 to 2012.
Since 2012, he has been with the Centre of Excellence in Telecommunications, School of Electrical and Information Engineering, the University of Sydney, Australia, where he is currently a Research Fellow. His research interests include channel and network coding, information theory, iterative multi-user detection, MIMO techniques and millimetre-wave communications. He has published over 40 peer-reviewed research papers in the leading international journals and conferences, and has served on a number of technical programs for international conferences such as ICC and WCNC.
\end{IEEEbiography}

\begin{IEEEbiography}[{\includegraphics[width=1in,height=1.25in,clip,keepaspectratio]{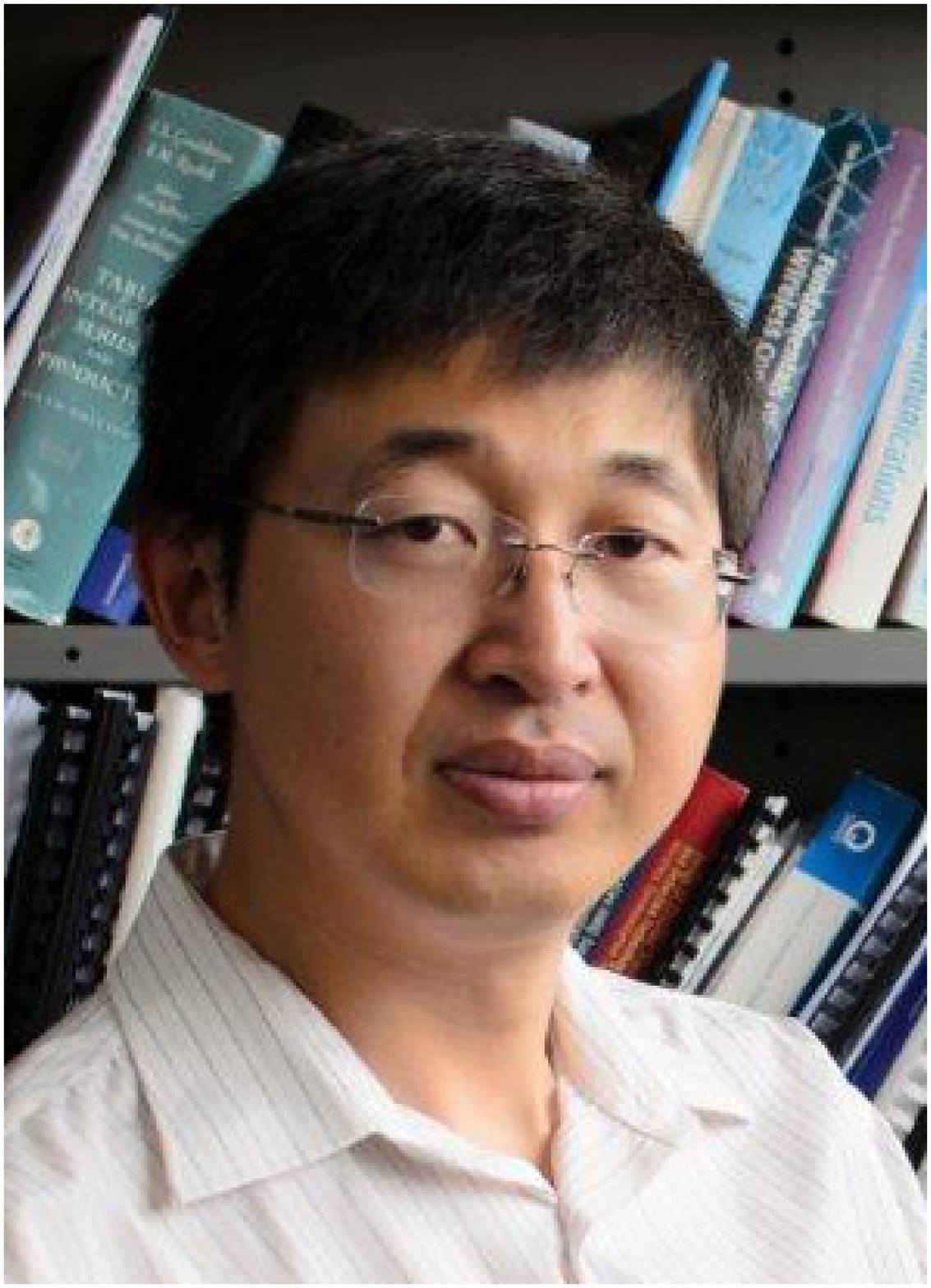}}]{Yonghui Li} (M'04-SM'09)
received his PhD degree in November 2002 from Beijing University of Aeronautics and Astronautics. From 1999 - 2003, he was affiliated with Linkair Communication Inc, where he held a position of project manager with responsibility for the design of physical layer solutions for the LAS-CDMA system. Since 2003, he has been with the Centre of Excellence in Telecommunications, the University of Sydney, Australia. He is now an Associate Professor in School of Electrical and Information Engineering, University of Sydney. He was the Australian Queen Elizabeth II Fellow and is currently the Australian Future Fellow. His current research interests are in the area of wireless communications, with a particular focus on MIMO, cooperative communications, coding techniques and wireless sensor networks. He holds a number of patents granted and pending in these fields. He is an executive editor for European Transactions on Telecommunications (ETT). He has also been involved in the technical committee of several international conferences, such as ICC, Globecom, etc.
\end{IEEEbiography}

\begin{IEEEbiography}[{\includegraphics[width=1in,height=1.25in,clip,keepaspectratio]{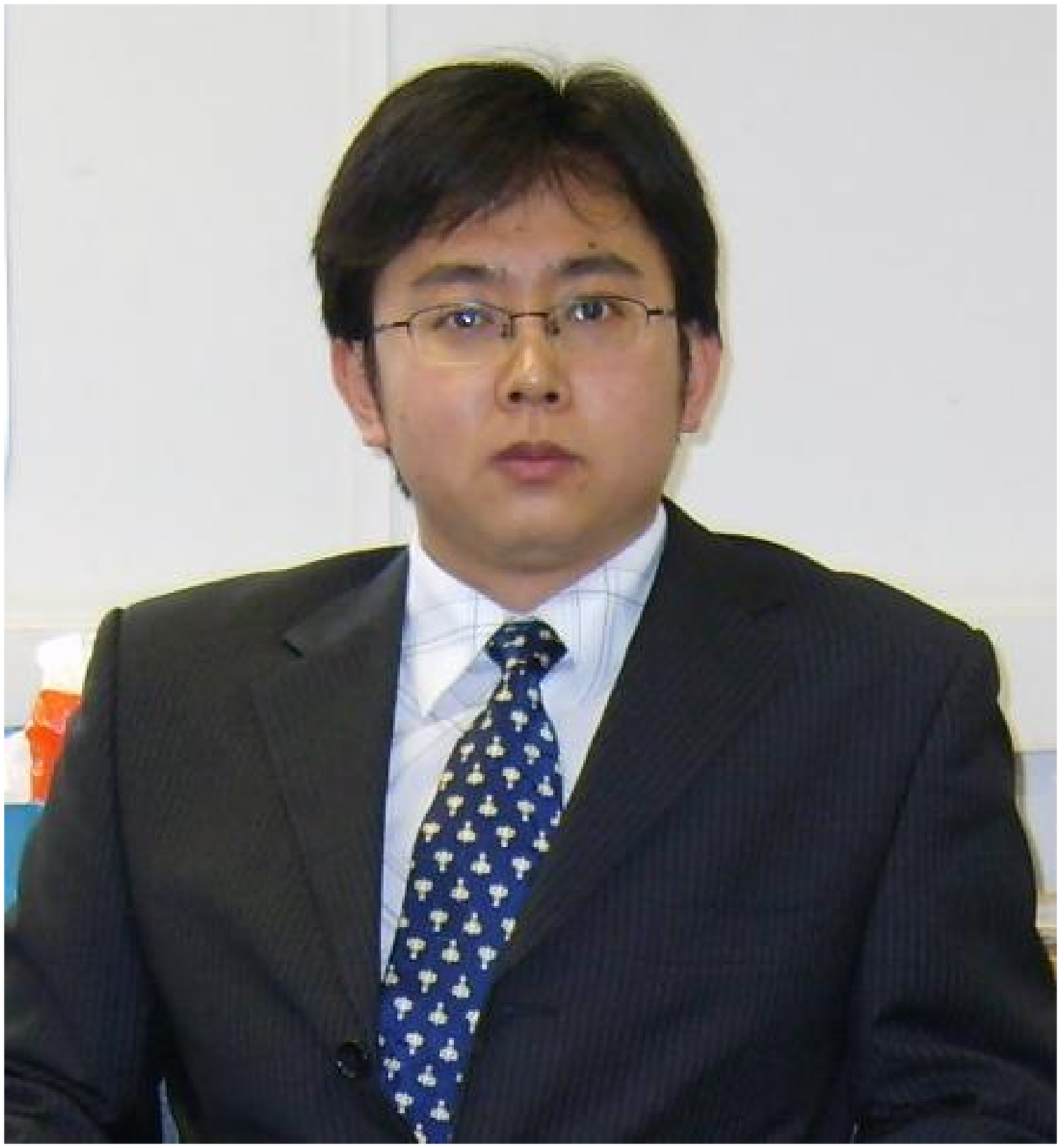}}]{Lingyang Song} (S'03-M'06-SM'12)
received his PhD from the University of York, UK, in 2007, where he received the K. M. Stott Prize for excellent research. He worked as a postdoctoral research fellow at the University of Oslo, Norway, and Harvard University, until rejoining Philips Research UK in March 2008. In May 2009, he joined the School of Electronics Engineering and Computer Science, Peking University, China, as a full professor. His main research interests include MIMO, OFDM, cooperative communications, cognitive radio, physical layer security, game theory, and wireless ad hoc/sensor networks.

He received the best paper awards in many conferences including IEEE International Conference on Wireless Communications, Networking and Mobile Computing (WiCOM 2007), the First IEEE International Conference on Communications in China (ICCC 2012), the 7th International Conference on Communications and Networking in China (ChinaCom2012), IEEE Wireless Communication and Networking Conference (WCNC2012), International Conference on Wireless Communications and Signal Processing(WCSP 2012) and IEEE International Conference on Communications (ICC2014). He is currently on the Editorial Board of IEEE Transactions on Wireless Communications, IET Communications and Journal of Network and Computer Applications. He is the recipient of 2012 IEEE Asia Pacific (AP) Young Researcher Award.
\end{IEEEbiography}

\begin{IEEEbiography}[{\includegraphics[width=1in,height=1.25in,clip,keepaspectratio]{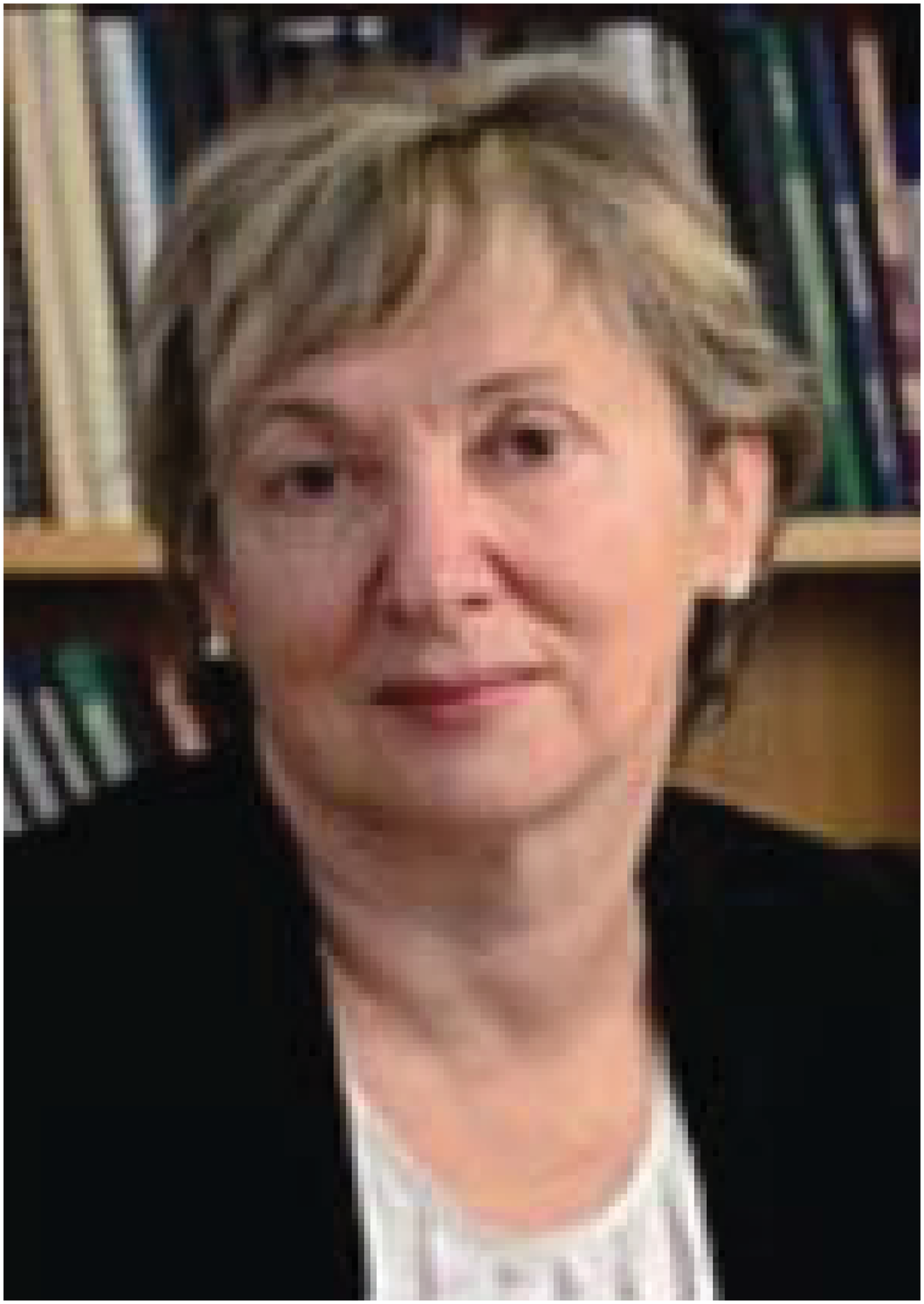}}]{Branka Vucetic} (F'03)
received the B.S.E.E., M.S.E.E., and Ph.D. degrees in electrical engineering, from the University of Belgrade, Belgrade, Yugoslavia, in 1972, 1978, and 1982, respectively. She currently holds the Peter Nicol Russel Chair of Telecommunications Engineering at the University of Sydney. During her career she has held various research and academic positions in Yugoslavia, Australia and UK. Her research interests include wireless communications, coding, digital communication theory and MIMO systems. She co-authored four books and more than three hundred papers in telecommunications journals and conference proceedings. She has been elected to the grade of IEEE Fellow for contributions to the theory and applications of channel coding.
\end{IEEEbiography}

%

%
\vfill
%
%




\end{document}